\documentclass[sigconf]{acmart} \usepackage{graphicx}
\usepackage{colortbl}
\usepackage[T1]{fontenc}
\usepackage{bm}
\usepackage{amsmath}
\usepackage{natbib}
\usepackage{amsfonts}
\usepackage[noend]{algorithmic}
\usepackage[nameinlink,capitalize]{cleveref}
\usepackage{float}
\usepackage{algorithm}
\usepackage{subfigure}
\usepackage{dsfont}
\usepackage{multirow}
\usepackage{array}
\usepackage{enumitem}
\usepackage{nccmath}
\setlist[itemize]{leftmargin=4mm}
\usepackage[colorinlistoftodos,textsize=tiny]{todonotes}
\usepackage{multibib}
\newcites{AP}{Appendix References}

\usepackage{soul}
\usepackage{xspace}
\definecolor{navy}{rgb}{0.1, 0.1, 0.8}
\definecolor[named]{gray}{rgb}{0.4, 0.4, 0.4}
\definecolor[named]{olive}{rgb}{0.1, 0.5, 0.1}
\definecolor[named]{ruby}{rgb}{0.8, 0.1, 0.3}
\definecolor{darkpastelgreen}{rgb}{0.01, 0.75, 0.24}
\definecolor{celestialblue}{rgb}{0.29, 0.59, 0.82}
\definecolor{coral}{rgb}{1.0, 0.5, 0.31}
\definecolor{Goldenrod}{rgb}{0.8,0.8,0}

\DeclareMathOperator{\E}{\mathbb{E}}
\DeclareMathOperator{\Prob}{\mathbb{P}}
\DeclareMathOperator{\Real}{\mathbb{R}}

\DeclareMathOperator{\His}{\mathcal{H}}
\DeclareMathOperator{\Cas}{\mathcal{C}}

\newcommand{\bracket}[1]{\left[#1\right]}

\newcommand{\numberthis}{\addtocounter{equation}{1}\tag{\theequation}}
\newcommand{\matr}[1]{\bm{#1}}

\copyrightyear{2023}
\acmYear{2023}
\setcopyright{acmlicensed}\acmConference[WWW '23]{Proceedings of the ACM Web Conference 2023}{May 1--5, 2023}{Austin, TX, USA}
\acmBooktitle{Proceedings of the ACM Web Conference 2023 (WWW '23), May 1--5, 2023, Austin, TX, USA}
\acmPrice{15.00}
\acmDOI{10.1145/3543507.3583481}
\acmISBN{978-1-4503-9416-1/23/04}

\begin{document}
\fancyhead{}

\newcommand{\titlename}{Interval-censored Transformer Hawkes: Detecting Information Operations using the Reaction of Social Systems}

\title{\titlename}

\author{Quyu Kong}
\affiliation{\institution{Alibaba Group \& \\ University of Technology Sydney}
  \city{Hangzhou}
  \country{China}
}
\email{kongquyu.kqy@alibaba-inc.com}

\author{Pio Calderon}
\affiliation{\institution{University of Technology Sydney}
  \city{Sydney}
  \country{Australia}
 }
\email{pio.calderon@student.uts.edu.au}

\author{Rohit Ram}
\affiliation{\institution{University of Technology Sydney}
  \city{Sydney}
  \country{Australia}
  }
\email{rohit.ram@student.uts.edu.au}

\author{Olga Boichak}
\affiliation{\institution{University of Sydney}
  \city{Sydney}
  \country{Australia}
  }
\email{olga.boichak@sydney.edu.au}

\author{Marian-Andrei Rizoiu}
\affiliation{\institution{University of Technology Sydney}
  \city{Sydney}
  \country{Australia}
}
\email{marian-andrei.rizoiu@uts.edu.au}

\newcommand{\activeRT}{\textsc{ActiveRT2017}\xspace}
\newcommand{\activepretrain}{\textsc{ActiveRT2016}\xspace}
\newcommand{\rncnix}{\textsc{RNCNIX}\xspace}
\newcommand{\io}{\textsc{IO}\xspace}

\begin{abstract}
Social media is being increasingly weaponized by state-backed actors to elicit reactions, push narratives and sway public opinion. These are known as Information Operations (IO). The covert nature of IO makes their detection difficult. This is further amplified by missing data due to the user and content removal and privacy requirements. This work advances the hypothesis that the very reactions that Information Operations seek to elicit within the target social systems can be used to detect them. We propose an Interval-censored Transformer Hawkes (IC-TH) architecture and a novel data encoding scheme to account for both observed and missing data. We derive a novel log-likelihood function that we deploy together with a contrastive learning procedure. We showcase the performance of IC-TH on three real-world Twitter datasets and two learning tasks: future popularity prediction and item category prediction. The latter is particularly significant. Using the retweeting timing and patterns solely, we can predict the category of YouTube videos, guess whether news publishers are reputable or controversial and, most importantly, identify state-backed IO agent accounts. Additional qualitative investigations uncover that the automatically discovered clusters of Russian-backed agents appear to coordinate their behavior, activating simultaneously to push specific narratives.
\end{abstract} 
\maketitle

\section{Introduction}

Online social platforms are explicitly designed to boost user interaction and engagement~\citep{wu2018beyond}.
Recently, there has been an increase in false and misleading claims being deliberately propagated and legitimized as tools of foreign interference~\citep{woolley2018introduction} --- also known as computational propaganda or Information Operations (IO).
IO actors leverage a wide spectrum of problematic online content (from satire and parody to manipulated and outright fabricated content) and behavior (inauthentic coordination) to elicit reactions and sway public opinion~\citep{wardle2017information}.
Content- and user-based detection tools are notoriously difficult to build due to language nuances and suffer from language drift and adversarial attacks.
Here, we start from the hypothesis that the actions of IO actors are designed to elicit particular reactions from the target audience.
This makes analyzing the posting behavior of users and the reaction of the social systems surrounding them a helpful starting point to map actors and agendas behind those posts.
This paper presents a novel approach to model the reactions of social systems, to disentangle the structural and functional types of users from the content posted by them.
Based solely on the timing of retweet cascades, we can separate YouTube videos' categories, quantify news publishers' trustworthiness, and even identify users involved in state-linked coordinated operations.

The ubiquitous diffusion cascades generated in the online environment, such as reshare events on Twitter and replies on Reddit, unveil the temporal dynamics of online items \citep{Zhao2015SEISMIC:Popularity,Mishra2016FeaturePrediction} and have been shown to be a proxy for characterizing them \citep{kong2020describing}.
A typical difficulty when analyzing reshare cascades is partial data loss --- i.e., when some of the events in the cascade are missing.
For instance, the \io dataset made available by the Twitter Moderation Research Consortium~\citep{tmrc2022} contains tweets from users relating to state-linked coordinated operations; however, for privacy reasons, it does not contain any of the activities of organic, non-malicious users.
Besides, a recent study by \citet{wu2020variation} comprehensively examined the sampling effect of retweet cascades collected from Twitter, where the loss of tweets may lead to modeling bias. 
We ask the question: \textbf{can we model reshare cascades with partial data loss, containing both event times and missing event counts?}
This work develops the Interval-censored Transformer Hawkes (IC-TH) model, an enhancement of the current state-of-the-art temporal point process model, the Transformer Hawkes model~\citep{zhang2020self}.
IC-TH leverages two sources of information --- the event timestamps and the counts of missing events.
We propose a unified data encoding scheme that jointly accounts for the two data sources, and we derive a novel log-likelihood function that augments the canonical point process likelihood function~\citep{Daley2008}.
Using synthetic experiments, we evaluate IC-TH robustness concerning data loss.
We find that cascades generated from distinct parameter combinations can be near-perfectly separated even when $90\%$ of the events in each cascade are missing.

Characterizing problematic content (such as mis- and dis-in\-for\-ma\-tion) or identifying malicious users typically demands significant manual labeling efforts.
In practice, only limited amounts of labeled data are available, raising the question:
\textbf{how do we leverage unlabeled cascade data for learning the representation of online items and users?}
Inspired by recent progress in pre-training deep neural nets~\citep{devlin2018bert,xu2022ccgl}, we propose a contrastive learning procedure to train IC-TH on unlabeled cascade groups. 
We build positive and negative pairs by sampling from cascade groups --- a cascade group contains all the cascades pertaining to a given online item or in which a given user participates.
We train the model to differentiate cascades from different groups and maximize the similarities among cascades within the same group.
Contrastive learning enables IC-TH to be pre-trained on large, unrelated datasets and later fine-tuned on the available labeled data.
We show that pre-training improves performance when only small amounts of labeled data are available.

The final question is \textbf{can we employ IC-TH to disentangle between content types and detect malicious online users?}
We test IC-TH on several tasks and case studies.
We study two tasks --- popularity prediction and online item categorization --- on three real-world cascade datasets (\activeRT, \rncnix and \io).
We show that IC-TH consistently outperforms the current state-of-the-art models, such as Mean Behavior Poisson~\citep{rizoiu2021interval}, HawkesN~\cite{kong2019modeling} and Transformer Hawkes~\citep{zuo2020transformer} for popularity prediction, and Dual Mixture Model~\citep{kong2020describing} and Transformer Hawkes~\citep{zuo2020transformer} for item classification.
Specifically, we augment the \io dataset with organic Twitter users and the prediction result on the \io dataset shows a macro-F1 of $98.7\%$ at separating users relating to Iran-, Russia- and Saudi Arabia-backed information operations and organic users.
Furthermore, embeddings generated by IC-TH allow for analyzing the strategies of groups of users. 
We identify three clusters of Russia-backed users; the users in each cluster appear to be coordinated in their behavior, activating simultaneously to push specific narratives such as the \textit{\#columbiachemicals} hoax in 2014.

The main contributions of this work are:
\begin{itemize}
    
    \item The interval-censored Transformer Hawkes for modeling the mixed temporal data formats of events and event counts, along with a unified temporal data coding scheme and a log-likelihood function.
    
    \item A pre-training procedure based on contrastive learning to leverage the large number of unlabeled cascade groups relating to different online items/users. 
    \item A set of experimental results on large real-world Twitter cascade datasets where we observe improved performances on item/user category and final popularity predictions from the proposed model and the pre-training procedure.
    \item Qualitative and quantitative analysis of user embeddings generated by our model on the \io dataset.
\end{itemize}
 \noindent \textbf{Related work.}
Diffusion cascades are commonly observed in the forms of individual events and event counts in between observation times. Event-based point process models employ different intensity functions (e.g., predefined parametric forms
\citep{hawkes1971spectra,Mishra2016FeaturePrediction,Zhao2015SEISMIC:Popularity,Rizoiu2017c,zhou2022},
deep neural nets~\citep{mei2017neural,Mishra2018ModelingPopularity,zuo2020transformer,zhang2020self,du2016recurrent}
and non-parametric variants~\citep{zhang2018efficient,zhou2013learning2}) to characterize the likelihood of event emergence, and the model parameters are typically estimated from the general point process likelihood function~\citep{Daley2008}.
On the other hand, count-based point process models generally acknowledge the loss in the interval-censored raw data and attempt to uncover the underlying parameters of the corresponding event-based models. 
For instance, \citet{kirchner2016hawkes} shows that the Hawkes process parameters can be obtained from event-count data using an integer-valued auto-regressive (INAR) model, and \citet{rizoiu2021interval} connect the Hawkes processes with a non-homogeneous Poisson process that can be fitted on event counts. 
This work extends both classes of models by proposing a Hawkes process variant based on the Transformer architecture with a novel likelihood function mixing both events and count data.

On the application front, prior works use the models above, including the Hawkes processes~\cite{kong2020describing} and deep learning models~\citep{sharma2020identifying,Ram2022} to encode online items and users based on their temporal activities for detecting coordinated accounts from Twitter, while earlier works apply pre-defined features to summarize online user behavior~\citep{addawood2019linguistic,zannettou2019disinformation}. Our work leverages a model pre-training procedure that further learns from a large unlabeled dataset to improve the derived embeddings of online items and users.
 \section{Preliminaries}\label{sec:preliminary}
This section covers the required preliminaries concerning modeling event cascades and series of interval-censored event counts.
We first mathematically define cascades in two distinct forms: event timestamps and event counts. 
We then introduce the Hawkes process~\citep{hawkes1971spectra} --- alongside its extensions HawkesN~\citep{Rizoiu2017c}, NeuralHawkes~\citep{mei2017neural} and Transformer Hawkes~\citep{zuo2020transformer} --- and Mean Behavior Poisson (MBP) process~\citep{rizoiu2021interval} for modeling event times and event counts, respectively.
As far as we are aware, no existing model can handle mixtures of event times and counts of missing events; 
therefore, a gap exists between models for interval-censored data and models for event cascades.

\textbf{Diffusion cascades.} 
On online social media platforms, such as Twitter, users can post/tweet content, and others may reshare or retweet it, resulting in cascades of retweet events~\citep{IJoC14866}. 
On other platforms, the sharing events may be only available as aggregated counts with selected granularity, such as Youtube, where only daily view counts are available~\citep{Wu2019}.
Formally, an event cascade, $\His_T=\{t_0, t_1, \ldots\}$, is a set of timestamps of individual sharing events observed until time $T$.
In the interval-censored setting, a set of observation times, $O_T = \{o_0, o_1, \ldots\}$, partitions an event cascade into time segments, and only the number of events $C(o_i, o_{i+1}]$ in each segment is observed.
Prior works successfully model diffusion cascades with both data forms~\citep{Zhao2015SEISMIC:Popularity,rizoiu2021interval}, as we discuss next.

\subsection{Models for event data}

\textbf{Hawkes process} is a particular type of point process model with the self-exciting property --- i.e., the future event intensity depends on all past events \citep{hawkes1971spectra}. 
Its event intensity at time $t$ is given as
\begin{align}\label{eq:hawkes}
    \lambda(t) = \mu + \sum_{t_i \in \His_T} \phi(t-t_i),
\end{align}
where $\phi: \Real^+ \rightarrow \Real^+$ is known as the decay function capturing the intensity excited by past events and $\mu$ is the background intensity which is normally considered $0$ for reshare cascades~\citep{Zhao2015SEISMIC:Popularity}. 
Common choices for the decay function include the exponential function, $\phi_{EXP} (\tau) = \kappa \theta e^{-\theta \tau}$, and the power-law function, $\phi_{PL} (\tau) = \kappa (\tau + c)^{-(1+\theta)}$~\citep{kong2019modeling}. 
A finite-population extension of Hawkes processes, dubbed \textbf{HawkesN processes}~\citep{Rizoiu2017c}, has shown superior modeling performance on online social media data~\citep{kong2019modeling}. 
The intensity function of HawkesN processes is modulated by the proportion of remaining unaffected individuals in the population, i.e.,
\begin{equation}
    \lambda(t) = \frac{N- N_t}{N} \sum_{t_i \in \His_T} \phi(t-t_i).
\end{equation}
where $N$ is the population size and $N_t$ is the number of events up to time $t$.

\textbf{Neural Hawkes process} generalizes the intensity function of vanilla Hawkes process with deep recurrent neural networks (RNNs)~\citep{mei2017neural,Mishra2018ModelingPopularity}, i.e.,
\begin{align}
    \lambda^N(t) = f(\matr{w}^\top \matr{h}(t)),
\end{align}
where $\matr{w}$ is a weight matrix and $\matr{h}(t)$ is a hidden state that encodes event history information produced by a continuous-time LSTM. $f(x) = \beta \log (1 + e^{\frac{x}{\beta}})$ is a softplus function parameterized by $\beta$.

\textbf{Transformer Hawkes (TH) process}~\citep{zuo2020transformer,zhang2020self} is the natural extension to the Neural Hawkes processes, leveraging an attention-based model~\citep{vaswani2017attention} that showed superior performance in sequence modeling compared to RNNs. 
The hidden state of Transformer Hawkes is obtained via a self-attention module. 
Specifically,
\begin{align}
    &\matr{h}(t_j) = \matr{H(j, :)},\\
    &\matr{H(j, :)}= ReLU(\matr{SW}_1 + \matr{b}_1)\matr{W}_2 + \matr{b}_2,  \\
    &\matr{S} = Concat(head_1, head_2, \cdots) \matr{W}^O, \\
    &head_i = Softmax\left(\frac{\matr{XW}^Q_i (\matr{XW}^K_i)^\top}{\sqrt{d_k}}\right)\matr{XW}^V_i,
\end{align}
where $\matr{W}^Q_i, \matr{W}^K_i \in \Real^{d_m \times d_k}$, $\matr{W}^V_i \in \Real^{d_m \times d_v}$, $\matr{W}^O \in \Real^{hd_v \times d_m}$, $\matr{W}_1 \in \Real^{d_m \times h}$, $\matr{b}_1 \in \Real^{h}$,$\matr{W}_2 \in \Real^{h \times d_m}$ and $\matr{b}_2 \in \Real^{d_m}$ are learned weights, $d_m$ is the embedding
dimension, $d_k$, $d_v$ are
the hidden dimensions of the projection subspace, and $h$ is the number of heads.
The input $\matr{X}$ is obtained via the temporal encoding, i.e.,
\begin{align}
    \matr{X(j, i)} =     \begin{cases}
        \cos (t_j / 1000^{\frac{i-1}{d_m}}) & \text{if $i$ is odd},\\
        \sin (t_j / 1000^{\frac{i}{d_m}}) & \text{if $i$ is even}.
      \end{cases}
\end{align}
where trigonometric functions are used to encode each event timestamp $t_j$ into a $d_m$-dimensional feature vector, $X(j,:)$.

\textbf{Parameter estimation.} 
The parameters of the models mentioned above are estimated via maximizing the general point process log-likelihood~\citep{Daley2008}, i.e.,
\begin{align}
    \mathscr{L}_{\mathrm{LL}} &= \sum_{t_i \in \His_T} \lambda(t_i) - \int_0^{T}\lambda(\tau) d\tau.
\end{align}

\subsection{Models for interval-censored data}
\label{subsec:mbp-notation}
\textbf{Mean Behavior Poisson (MBP)} processes are a particular variant of Poisson processes that fit on interval-censored data.
MBP is defined as the point process whose event intensity is the mean Hawkes process intensity over all the possible realizations of a parameter set.
As a result, MBP's parameters directly correspond to their Hawkes process counterparts~\citep{Calderon2021,rizoiu2021interval}.
Specifically, given a Hawkes process in \cref{eq:hawkes}, its MBP equivalence is defined by a deterministic intensity function
\begin{equation}
    \xi(t) = \mu + \int_0^{t} \xi(\tau) \phi(t-\tau) d\tau,
\end{equation}
where $\xi(t) = \E_{\His_t} [\lambda(t)]$.
The MBP log-likelihood is
\begin{align}\label{eq:icll}
    \mathscr{L}_{\mathrm{IC}-\mathrm{LL}}(\theta) &= \sum_{i=1}^{m} \mathrm{C} \left( o_{i-1}, o_i \right) \log \Xi\left(o_{i-1}, o_{i}; \theta \right) - \sum_{i=1}^{m} \Xi\left(o_{i-1}, o_{i} ; \theta \right)
\end{align}
where $\mathrm{C} \left(o_{i-1}, o_i \right)$ is the number of events recorded in the interval $\left(o_{i-1}, o_i \right)$, and we define the compensator
\begin{align}
\Xi\left(o_{i-1}, o_{i}; \theta \right) = \int_{o_{i-1}}^{o_i} \xi (z) dz
\end{align}

\begin{figure}[!tbp]
    \centering
    \includegraphics[width=0.47\textwidth]{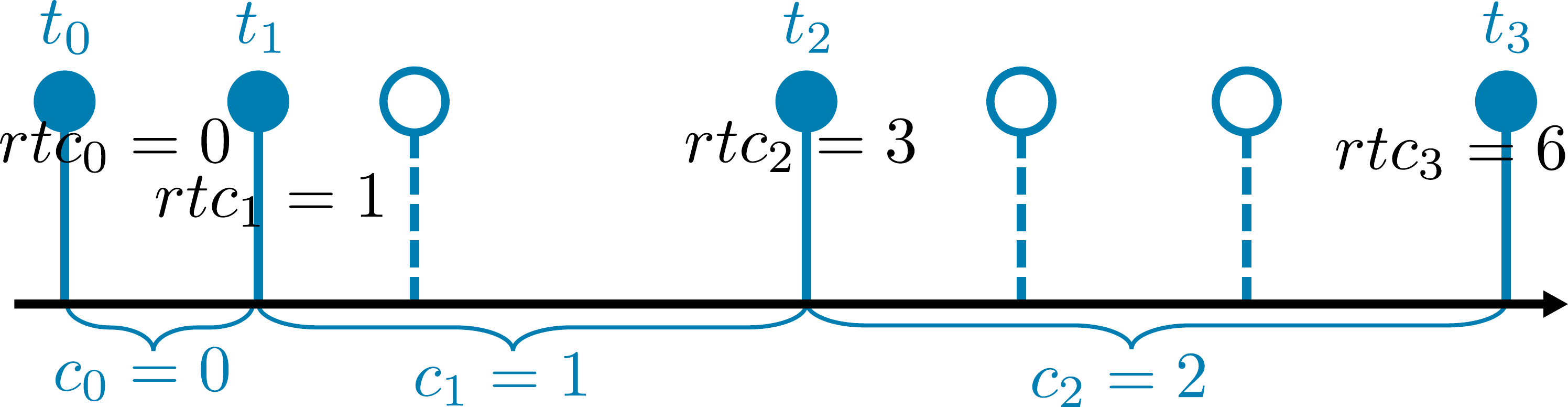}
	\caption{
		$4$ observed events (lollipops with solid lines) of an example retweet cascade with the presence of unobserved retweet events (lollipops with dotted lines). Only retweet counts are given (i.e., $rtc_0, rtc_1, rtc_2,\cdots$) by Twitter API and the exact timestamps of unobserved events are missing.
	}
	\label{fig:teaser}
\end{figure}
 \section{Interval-censored Transformer Hawkes} 
This section first explores where the mixing of event counts and event times is applicable (\cref{subsec:missing_events}). 
We then propose to encode such data in a flexible format, extend the Transformer Hawkes to model it, and propose a novel log-likelihood function (\cref{subsec:ic-th}). 
Finally, we discuss a contrastive learning and pre-training procedure for training IC-TH without large volumes of labeled data (\cref{subsec:pre-training}).

\subsection{Missing events on Twitter}
\label{subsec:missing_events}

\textbf{Twitter sampling in data collection.}
While existing studies modeling information diffusions on Twitter generally analyze diffusion cascades with individual retweet events crawled via the official API endpoint~\citep{Zhao2015SEISMIC:Popularity,Mishra2016FeaturePrediction,kong2019modeling}, the sampled-down effect from the API~\citep{wu2020variation} has been usually overlooked and may lead to modeling bias. For instance, the sampled streaming API returns roughly 1\% of all public tweets in real-time and the widely adopted filtered streaming API also rate limits on popular search queries.

\textbf{Piecing together the missing events count.}
The missing event counts between two consecutive observed events can be derived by leveraging the `$retweeted\_count$' in tweet metadata.
This field is included in every retweet and indicates how many times the original tweet has been retweeted (including by the current retweet).
\cref{fig:teaser} depicts an example Twitter cascade with missing tweets as dashed lollipops.
Four tweets are observed at times $t_0$, $t_1$, $t_2$ and $t_3$.
$rtc_i$ indicates the `$retweeted\_count$' value in the $i$-th observed tweet.
We compute $c_0, c_1, c_2, \cdots$ as the number of \emph{missing} retweets between two observed tweets. 
This leads to a cascade that mixes both event counts and individual events.

\textbf{A mixed data format.}
We propose the triple representation $(o_i, d_i, c_i)$ for the $i^{th}$ observation, where $o_i$ is the starting observation time, $d_i$ is the observation duration and $c_i$ is the number of events (observed or missing) within the time segment $(o_i, o_i + d_i)$ --- note, excluding interval ends. 
Using this notation, we denote an observed event at $t_i$ as $(t_i-dt, 2 dt, 1)$ where $dt$ is an infinitesimal time period.
We denote the missing events between $t_i$ and $t_{i+1}$ as $(t_i, t_{i+1}-t_i, c_i)$.
Note that $c_i$ would correspond to $\mathrm{C}(t_i, t_{i+1})$ in MBP's interval-censored notation in \cref{subsec:mbp-notation}.
Finally, we can denote the reshare cascade as $\His^* = \{(t_i, d_i, c_i) \mid i=0,\cdots,m, o_i+d_i \leq T\}$ given a maximum observation time $T$ and $m$ the number of observed timestamps. For instance, one can denote the cascade in \cref{fig:teaser} as $= \{(t_0-dt, 2dt, 1), (t_0, t_1-t_0, 0), (t_1-dt, 2dt, 1), (t_1,t_2-t_1, 1), \cdots \}$.

\subsection{Interval-censored Transformer Hawkes}
\label{subsec:ic-th}
In this section, we propose the Interval-censored Transformer Hawkes (IC-TH) that generalizes the Transformer Hawkes accounting for event times and missing event counts. 
IC-TH takes data in the interval-censored event sequences format proposed in \cref{subsec:missing_events}, and produces the hidden states using the self-attention mechanism proposed by \citet{zuo2020transformer}.

\textbf{Model event intensity.}
Due to missing event timestamps (only their counts are available),
IC-TH models reshare cascades using the expected conditional event intensity.
We adopt the notations of \citet{rizoiu2021interval} and denote the expected event intensity function as
\begin{align}
    \xi(t \mid \His^*) = \E_{\His^*_u} \bracket{\lambda(t)} = f(\matr{w}^\top \matr{h}(t)),
\end{align}
where $\His^*_u=\{(t_i, d_i, c_i) \in \His^* \mid d_i > 2dt, c_i > 0 \} \subseteq \His^*$ is the subset of triples denoting missing event counts.
The intensity function is modeled from the hidden state $\matr{h}(t)$ via a softplus function $f(\cdot)$ and a weight matrix $\matr{w}$.

We note that when $\His^*_u = \emptyset$ or $\His^* = \His^*_u$, $\xi(t \mid \His^*)$ degrades to $\lambda(t \mid \His^*)$ or $\xi(t)$ discussed in \cref{sec:preliminary}, respectively.

\textbf{Encoding $\His^*$.}
IC-TH employs the same temporal encoding procedure as TH to convert the timestamp $t_i$ into an input vector $X_i'$.
However, unlike the timestamps, the duration and event count information needs extra encoding layers before being fed to the self-attention module. 
Inspired by \citet{li2017time}, we encode the duration $d_i$ and the event count $c_i$ as masks to $X_i'$. 
Specifically, for $d_i$, we compute the context vector $c^d_i$ using a fully connected layer $f_{\theta}$, i.e., $c^d_i = f_{\theta}(\log (d_i))$. 
The duration mask $m^d_i$ is then obtained via a linear transformation parameterized by $\matr{W}_d$, $\matr{b}_d$ and a sigmoid function $\sigma$, i.e., $m^d_i = \sigma (c^d_i \matr{W}_d + \matr{b}_d)$. 
Similarly, the event count mask $m^c_i$ is computed following the same layers with a different parameter set. 
Finally, $X_i = X_i' \bigodot m^d_i \bigodot m^c_i$, where $\bigodot$ is an element-wise multiplication.

\textbf{Log-likelihood function.} Starting from \cref{eq:icll}, we compute the log-likelihood function for the mixings of event counts and individual events.
\begin{align*}
    &\mathscr{L}_{\text {IC-TH-LL }}(\theta) = \sum_{i \in \His^*} c_{i} \log \Xi\left(o_{i}, o_{i}+d_{i}\right) - \sum_{i \in \His^*} \Xi\left(o_{i}, o_{i}+d_{i}\right) \\
    &= \underbrace{\sum_{i \in \His^*_u} c_{i} \log \Xi\left(t_{i}, t_{i+1}\right)}_{\text{missing event counts}} + \underbrace{\sum_{i \in \His^*_c} \log \Xi\left(t_{i} - dt, t_{i}+d t\right)}_{\text{observed event times}} \\
    & \hspace{0.5cm} - \sum_{i \in \His^*} \Xi\left(t_{i}, t_{i+1}\right) \\
    &=\sum_{i \in \His^*_u} c_{i} \log \Xi\left(t_{i}, t_{i+1}\right)+\sum_{i \in \His^*_c} \log \xi\left(t_{i}\right) 2 d t - \sum_{i \in \His^*} \Xi\left(t_{i}, t_{i+1}\right) \\
    &=\sum_{i \in \His^*_u} c_{i} \log \Xi\left(t_{i}, t_{i+1}\right)+\sum_{i \in \His^*_c} \log \xi\left(t_{i}\right) - \sum_{i \in \His^*} \Xi\left(t_{i}, t_{i+1}\right)  \\
    &\hspace{0.5cm}+ \sum_{i \in \His^*_c}\log 2 d t, \numberthis
\end{align*}
where $\His^*_c = \{(t_i, d_i, c_i) \in \His^* \mid d_i = 2dt, c_i =1 \}$ is the subset of triples denoting observed event times. 
This is equivalent to
\begin{align}
    &\mathscr{L}_{\text {IC-TH-LL }}(\theta) \nonumber \\
    &= \sum_{i \in \His^*_u} c_{i} \log \Xi\left(t_{i}, t_{i+1}\right)+\sum_{i \in \His^*_c} \log \xi\left(t_{i}\right) - \sum_{i \in \His^*} \Xi\left(t_{i}, t_{i+1}\right).
\end{align}

\textbf{The Linformer trick.} 
The quadratic complexity of the transformer architecture in memory restricts the encoding of long cascades and large cascade groups. 
Therefore, we employ the Linformer trick, which optimizes the model to linear complexity~\citep{wang2020linformer}. 
Specifically, the $head$s in Linformer are computed as
\begin{align}
    head_i = Softmax\left(\frac{\matr{XW}^Q_i (\matr{E_i XW}^K_i)^\top}{\sqrt{d_k}}\right)\matr{F_i XW}^V_i
\end{align}
where $E_i, F_i \in \Real^{k\times n}$ are two projection matrices that simplify the computation via $k \ll n$.

\textbf{Implementation details.} 
Throughout the paper, we conduct our experiments on a machine with an Intel Xeon CPU @ 2.20GHz processor and with Nvidia Tesla Volta V100 GPUs (32 GB memory). 
The proposed model is implemented in PyTorch 1.9.0 \citep{NEURIPS2019_9015}.

\subsection{Model pre-training}
\label{subsec:pre-training}

\textbf{Intuition.}
Here, we build a contrastive representation learning pipeline to obtain informative embeddings from groups of unlabeled diffusion cascades.
The intuition is to leverage large amounts of unlabeled (and possibly unrelated to the learning task) diffusion cascades from the Twitter API.

We group the cascades into discrete groups based on a specific criterion --- such as all cascades in which a given user participates or cascades related to a given online item (such as a YouTube video).
We build the representation for a group by aggregating the cascade embeddings for cascades of that group.
The contrastive representation learning algorithm aims to maximize latent distances between a group representation and other groups and minimize its latent distance to its own cascades.

\textbf{Technical detail.}
We first augment the dataset by building positive and negative pairs of cascade groups via downsampling. Specifically, given a group of cascades, $\Cas_i = \{\His_1, \His_2, \cdots\}$ posted by the same user or related to the same online item denoted as $i$, we randomly split the cascade group into two, $\Cas_{i,1}$ and $\Cas_{i,2}$, and consider them as a positive pair.
Naturally, examples formed by pairing cascade groups related to different items or users are deemed negative.

We then obtain the cascade embeddings $h_{i,1}$ and $h_{i,2}$ with the proposed model. 
Following prior works \citep{xu2022ccgl}, we further introduce a MLP-based projection head to a new representation, i.e., $z_{i,j} = MLP(h_{i,j})$.
Finally, we optimize the contrastive loss function \citep{chen2020simple}:
\begin{align}
    \mathscr{L}^{contrastive}_{i} = -\log \frac{exp(f_{sim}(z_{i,1}, z_{i,2})/\tau)}{\sum_{k=1}^{N} \sum_{l=1}^2\sum_{j=1}^{2} exp(f_{sim}(z_{i,l}, z_{k,j})/\tau)},
\end{align}
where $f_{sim}(z_1, z_2)$ is the cosine similarity between $z_1$ and $z_2$. \begin{table}[tbp]
	\centering
	\setlength{\tabcolsep}{4pt}
	\caption{Statistics of the datasets.}
\resizebox{\linewidth}{!}{
	\begin{tabular}{rrrrr}
	  \toprule
& \activepretrain & \activeRT & \rncnix & \io \\
	  \midrule
\#items & $155,105,987$ videos & $75,717$ videos & $102,429$ articles & $32,486$ users \\
\#cascades & $881,587,021$ & $30,535,891$ & $8,129,126$ & $19,476,766$ \\
      \#tweets & $1,212,945,195$ & $85,334,424$ & $56,397,252$ & $22,845,053$ \\
\bottomrule
	\end{tabular}
    }
	\label{tab:dataset-profiling}
\end{table}
 \begin{figure*}[!tbp]
    \centering
    \newcommand\myheight{0.14}
    \subfigure[$\Prob_m=0$]{
        \includegraphics[height=\myheight\textheight,page=1]{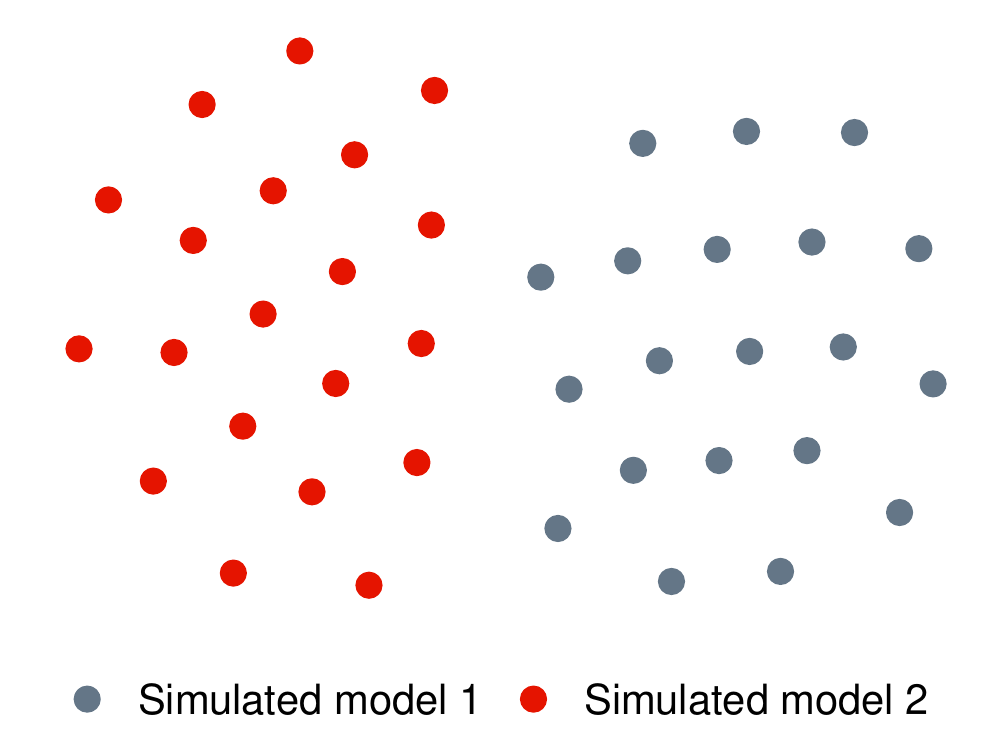}
        \label{fig:simulation_a}
    }
    \subfigure[$\Prob_m=0.5$]{
        \includegraphics[height=\myheight\textheight,page=2]{images/simulations.pdf}
        \label{fig:simulation_b}
    }
    \subfigure[$\Prob_m=0.8$]{
        \includegraphics[height=\myheight\textheight,page=3]{images/simulations.pdf}
        \label{fig:simulation_c}
    }
    \subfigure[$\Prob_m=0.9$]{
        \includegraphics[height=\myheight\textheight,page=4]{images/simulations.pdf}
        \label{fig:simulation_d}
    }
	\caption{
        t-SNE~\citep{maaten2008visualizing} visualizations of the group embeddings of synthetic cascades obtained using IC-TH.
        Each circle represents a group of cascades.
        The events in each cascade are removed with the probability $\Prob_m$, varying from no missing data ($\Prob_m = 0$, (a)) to $90\%$ of the events missing ($\Prob_m = 0.9$, (d)).
        Apart from down-sampling, the groups and the cascades are identical across (a)-(d).
	}
	\label{fig:simulation}
\end{figure*}
 
\section{Experiments and results} \label{sec:experiment}

In this section, we present the quantitative performance results for IC-TH.
First, we present the four datasets and learning tasks (\cref{subsec:datasets}).
Next, we study the robustness of IC-TH concerning data loss (\cref{subsec:data-loss}).
Finally, we evaluate IC-TH against the state-of-the-art in two predictive tasks: popularity prediction (\cref{subsec:popularity-prediction}) and online item categorization (\cref{subsec:category-prediction}).

\subsection{Datasets}
\label{subsec:datasets}
This work employs four real-world Twitter datasets. 
Two datasets originate from prior works (\activeRT and \rncnix).
We contribute two new datasets:
one for model pre-training (\activepretrain) and an Information Operations (\io) dataset based on the unhashed datasets provided by Twitter Moderation Research Consortium~\citep{tmrc2022}.

\textbf{\activeRT}~\citep{kong2020describing} was constructed by \citet{kong2020describing} by collecting tweets mentioning popular Youtube videos. 
Cascade groups are constructed based on the YouTube video ID: all the cascades mentioning the same video will be placed in the same cascade group.
The categorization task for this dataset is predicting the video's category (Entertainment, Gaming, Music and News\&Politics, etc.) based on retweet cascade dynamics.
This dataset suffers from Twitter API's down-sampling, given the popularity of YouTube videos on Twitter (see \citet{wu2020variation}).
We augment each retweet cascade with the missing event counts between consecutively recorded retweets (as described in \cref{subsec:missing_events}).  

\textbf{\activepretrain.} 
We contribute a dataset by following the same setup as \activeRT using tweets from 2016. 
We employ \activepretrain to pretrain the IC-TH to obtain a better representation of cascade groups. 
As we use the Twitter API v2 for collecting this dataset, the cascades are complete. 
To make \activepretrain comparable with \activeRT, we sample cascades at the same level as \activeRT.

\textbf{\rncnix~\citep{bruns2020news}.} 
The tweets in this dataset were collected 
by querying the Twitter search endpoint for tweets mentioning articles from a list of controversial news publishers and a list of leading Australian news outlets~\citep{bruns2016big,bruns2016publics,bruns2020news}.
The cascade groups were constructed based on the news publishers, i.e., all cascades relating to news articles from the same publisher are in the same group.
The categorization task for this dataset is predicting whether a news publisher is reputable or controversial based on retweet cascade dynamics.
The data is obtained retrospectively using a paid service, and the cascades are complete without the sampling-down effect.

\textbf{Information Operations (\io) dataset}
contains the users (also known as IO operatives) and their tweets linked to state-backed information operations identified and released officially by Twitter Moderation Research Consortium~\citep{tmrc2022}.
We obtained the unhashed version of this dataset in Nov 2021;
the dataset spans from Nov 4, 2010, to Aug 21, 2020.
We select IO operatives from three countries (Russia, Iran and Saudi Arabia) with the most significant number of users and highest activity.

We augment the dataset with a matched set of organic Twitter users, i.e., users who were involved in the discussions with the IO operatives but were not identified by Twitter as operatives.
Building the organic set presents challenges, as when Twitter suspends an IO operative account, it also deletes all its activity, including the retweets emitted by organic users.
We cannot use the Twitter API to search for the activity of IO operatives (as all of their traces have been removed), and the dataset does not contain organic activity.
Instead, we cross-link with a 1\% sampled tweet stream. 
We follow the following steps:
\begin{enumerate}
    \item We first collect the IDs of tweets posted or retweeted by removed users.
    \item We augment the tweets by searching for the retweet cascades using these tweet IDs in a complete archive of $1\%$ sampled tweet stream\footnote{https://developer.twitter.com/en/docs/twitter-api/tweets/volume-streams/introduction} hosted at \url{archive.org}\footnote{https://archive.org/details/twitterstream}.
    \item We consider users who participated in these retweet cascades but were not removed by Twitter as organic users.
To obtain users with similar popularity levels, we sample $10,000$ organic users by matching the follower count distribution of removed users.
   \item  Last, we repeat the first two steps for organic users to augment the dataset with all retweets cascades engaged by them.
\end{enumerate}
We construct cascade groups based on the users, i.e., all the cascades in which a given user participates are placed in the same group.
The categorization task is predicting whether a user is an IO operative from one of the three countries or an organic user (4 classes).
\cref{tab:dataset-profiling} summarizes the basic statistics of the four datasets.

\subsection{Effect of data loss on IC-TH}
\label{subsec:data-loss}

In this section, we evaluate the robustness of IC-TH's learned group embeddings concerning missing data.
We deploy a synthetic setup in which cascades are synthetically generated and grouped into cascade groups.
Events are randomly removed with given probabilities.
Finally, we generate group embeddings using the contrastive learning loss and evaluate the separability based on the group type.

\textbf{Setup.}
We sample the cascades from two Hawkes process models --- a power-law decay kernel and an exponential decay kernel, with randomly selected parameters. 
We sample $10,000$ cascades for each model, split into $20$ groups of $500$ cascades.
This results in $40$ groups of cascades. 
We synthetically emulate Twitter's sample-down mechanism for each cascade (see \citet{wu2020variation} for an in-depth discussion of Twitter's sample-down mechanism).
We choose a sampling-down probability $\Prob_m$, and we remove the tweets in each cascade independently and randomly with the probability $\Prob_m$.
Once an event is marked as missing, we remove it from its cascade and update the event count in the time interval. 
We conduct contrastive learning on the simulated cascade groups and apply a widely employed dimension reduction tool, t-SNE~\citep{maaten2008visualizing}, on the group embeddings for visualization. 
We repeat the experiments with varying $\Prob_m$ missing probability to examine the effect of different levels of data loss.

\cref{fig:simulation} depicts the embeddings of cascade groups produced from the trained IC-TH, for the various $\Prob_m$.
When there is no missing data ($\Prob_m = 0$) or half of the tweets are missing ($\Prob_m = 0.5$), the color groups are perfectly separated.
As the data loss increases ($\Prob_m > 0.5$), the color groups start mixing together.
However, we achieve near-perfect separability even at very large data sampling rates ($\Prob_m = 0.9$ in \cref{fig:simulation_d}).
This shows that IC-TH captures the differences in cascade groups even as most events are missing and represented as missing event counts.

\begin{figure}[!tbp]
    \centering
    \includegraphics[width=0.47\textwidth]{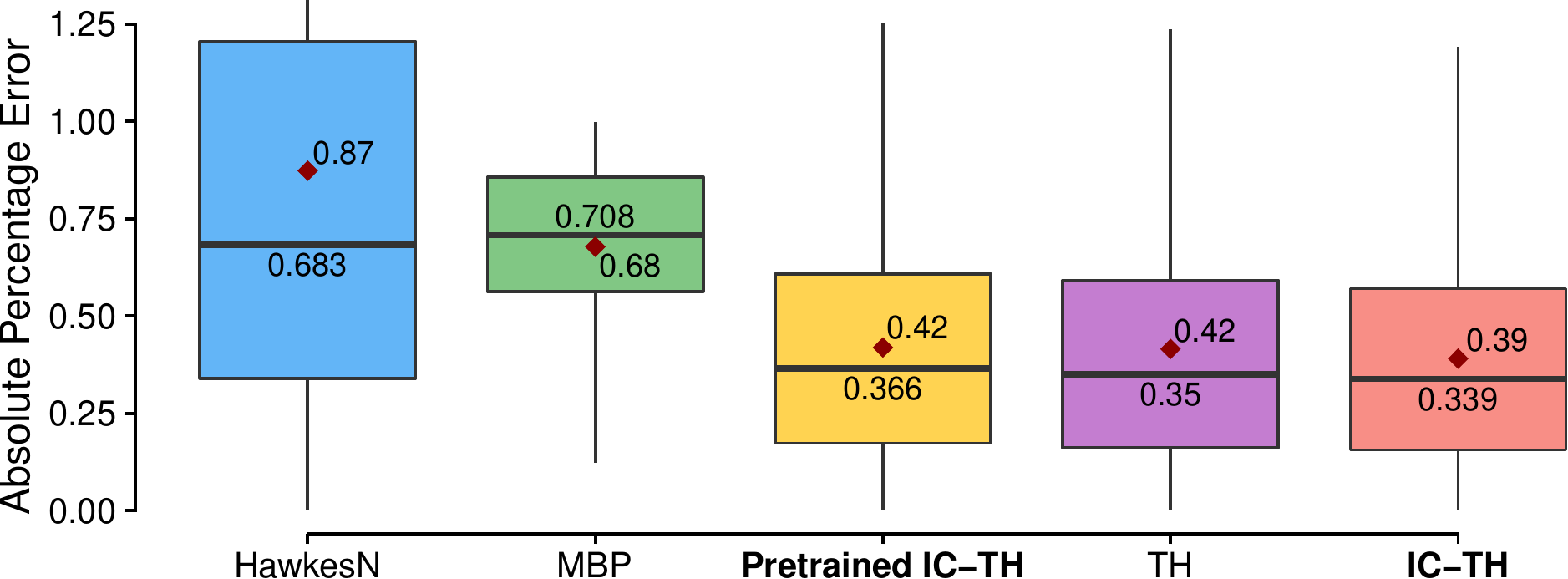}
	\caption{
        Performance of predicting final popularity measured using the absolute percentage error (APE) on \activeRT --- lower is better.
        We compare IC-TH (with and without pre-training, in \textbf{bold font}) against three state-of-the-art baselines: HawkesN~\cite{kong2019modeling}, MBP~\citep{rizoiu2021interval} and Transformer Hawkes (TH)~\citep{zuo2020transformer}.
        The boxplots summarize all the cascades in the dataset.
        The red diamond shows the mean APE.
	}
	\vspace{-5mm}
	\label{fig:pop_pred}
\end{figure}
 \subsection{Future popularity prediction}
\label{subsec:popularity-prediction}
In this section, we evaluate IC-TH for the cascade future popularity prediction task on the \activeRT dataset.
We evaluate the impact of pre-training for this task by testing IC-TH with and without pre-trained weights on the \activepretrain dataset.

\textbf{Setup and baselines.}
We compare IC-TH against the baseline models discussed in \cref{sec:preliminary}.
This includes an event count model (MBP~\citep{rizoiu2021interval}) and two event time models (the generalized HawkesN~\cite{kong2019modeling} and Transformer Hawkes~\citep{zuo2020transformer}). 
As the \activeRT is a sampled-down dataset, it contains both retweet counts and retweet events.
We process the data to adapt it to the requirements of each baseline as follows.
For event time models, we model only the observed event times in a cascade. 
For the event count model (i.e., MBP), we consider the observed event timestamps as observation times and compute the total event counts for each observation period --- e.g., $c_1+1$ for the observation period $t_0$ to $t_1$. 
These operations introduce information loss; however, we highlight that we are unaware of any other model that can leverage event count and individual event data simultaneously. 

\textbf{Results.}
\cref{fig:pop_pred} shows the results of the popularity prediction.
Visibly, all transformer-based approaches (IC-TH and TH) provide a significant performance boost over the generative approaches (HawkesN and MBP).
Furthermore, our proposed IC-TH outperforms TH in terms of both mean and median Absolute Percentage Error (APE).
Surprisingly, unlike the categorization prediction (see next section), pre-training does not boost performance.
The pre-trained IC-TH shows performances on par with TH. 

\begin{table}[tbp]
	\centering
	\caption{
		Macro-F1 of predicting the category of YouTube videos (on \activeRT), controversial news publishers (\rncnix) and state-backed Information Operations operatives (\io).
		We compare three flavors of IC-TH against two state-of-the-art models.
		Higher is better.
	}
\begin{tabular}{lrrr}
		\toprule
		Models & \activeRT & \rncnix & \io \\
		\midrule
		DMM~\citep{kong2020describing} & 0.488 & 0.675 & 0.968 \\
		TH~\citep{zuo2020transformer} & 0.469 & 0.823 & 0.983 \\
		IC-TH w/o missing counts & 0.495 & 0.840 & 0.985\\
		IC-TH & 0.499 & - & 0.987 \\
		Pre-trained IC-TH & 0.503 & 0.853 & 0.987  \\
		\bottomrule
	\end{tabular}
\label{tab:categorical_prediction}
	\vspace{-2mm}
\end{table} 
\begin{figure}[!tbp]
    \centering
    \includegraphics[width=0.47\textwidth]{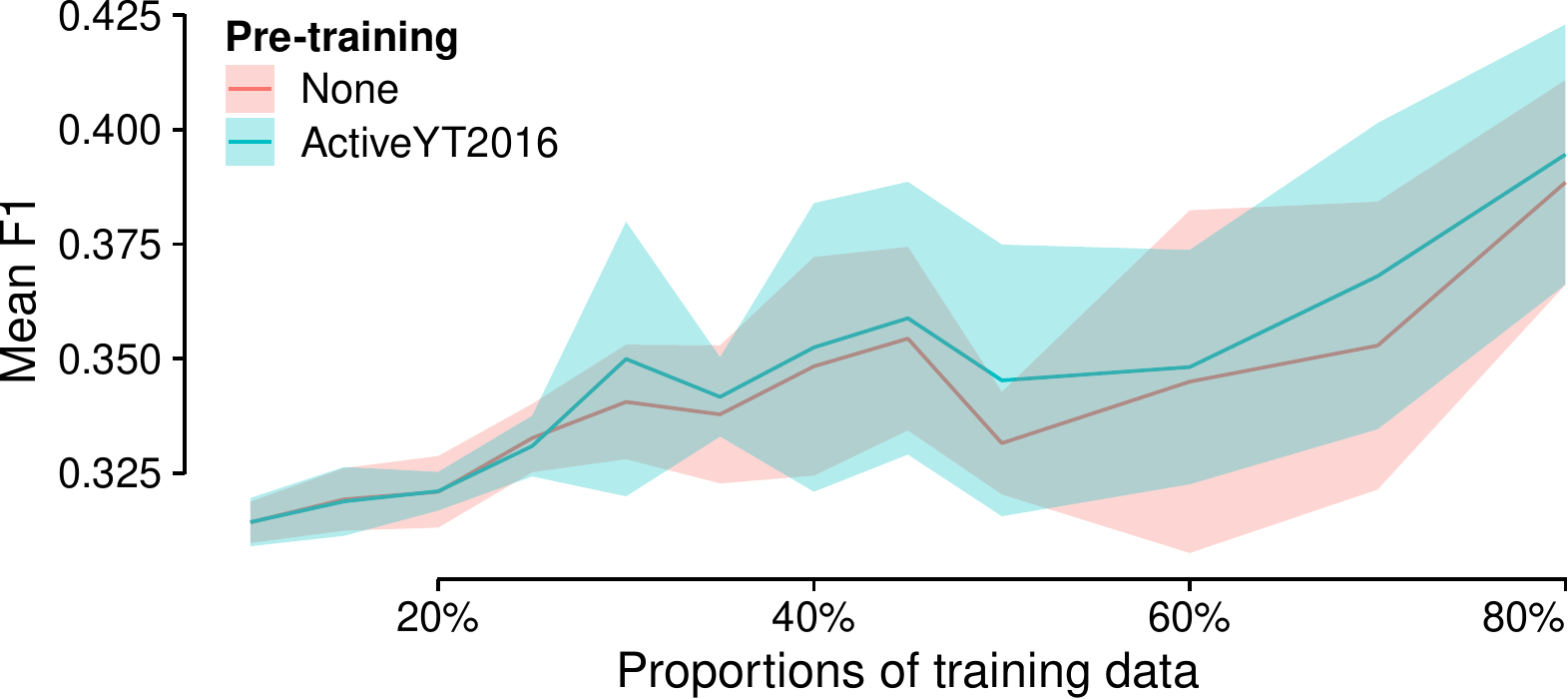}
	\caption{
		Comparison of prediction performances on \activeRT of IC-TH with and without pre-trained weights after being trained on different proportions of training data. Mean F1 scores and standard deviations among $10$ runs are shown as colored lines and areas, respectively.
	}
	\vspace{-5mm}
	\label{fig:f1_prop_train}
\end{figure}
 \begin{figure*}[!tbp]
    \centering
    \newcommand\myheight{0.22}
    \subfigure[]{
        \includegraphics[height=\myheight\textheight]{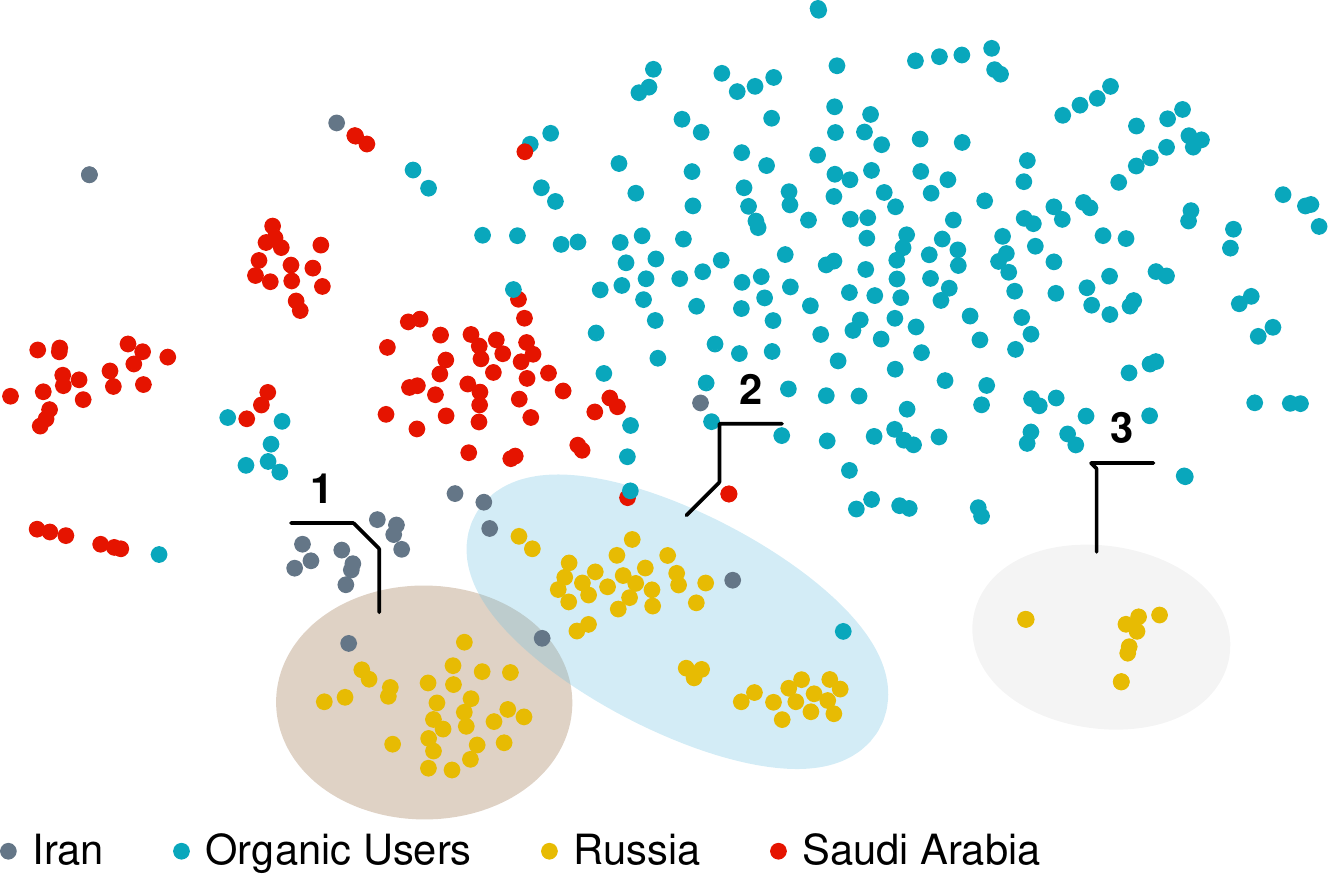}
        \label{fig:twitter_takedown}
    }\subfigure[]{
        \includegraphics[height=\myheight\textheight]{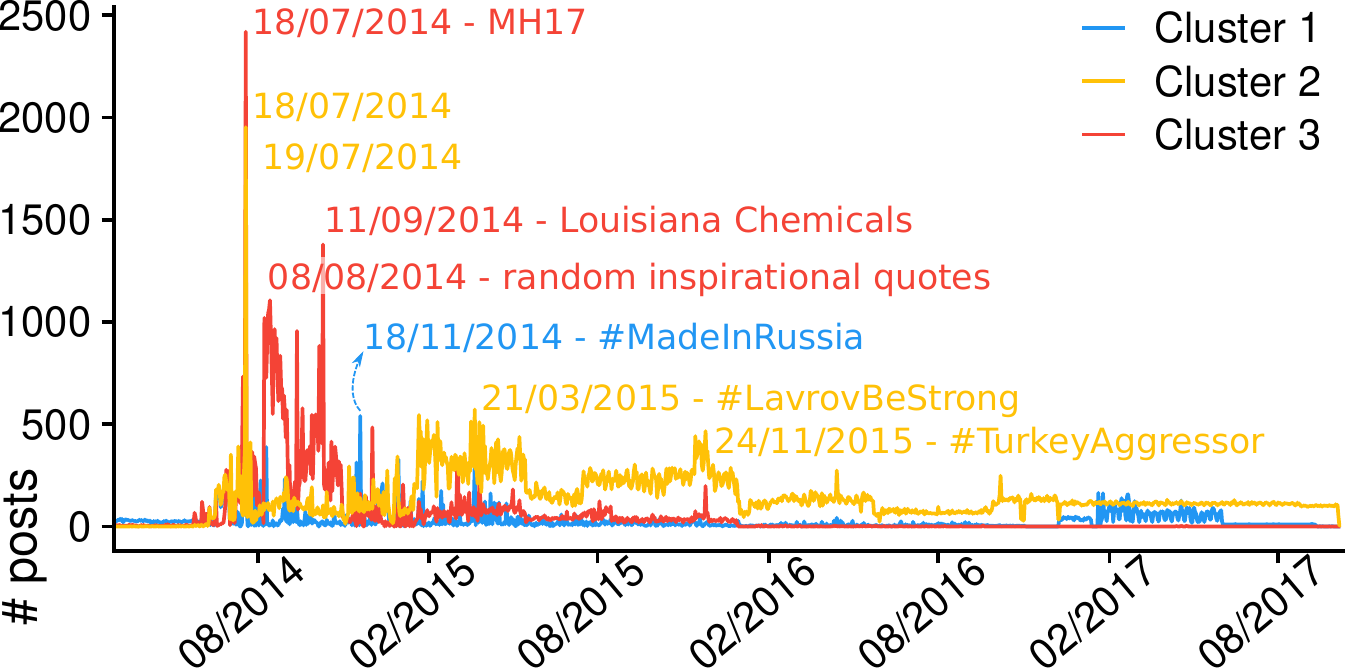}
        \label{fig:russia_op}
    }\vspace{-5mm}
	\caption{
        Understanding the strategies of state-backed agents.
		(a) T-SNE visualizations of the users in the \io dataset.
        The colors indicate the countries of origin for state-backed agents, as identified by Twitter.
        Three clusters of Russian-backed users are highlighted.
        (b) Temporal patterns of coordinated activity across the clusters of Russian-backed users.
        Annotations show the most used messages and hashtags on particular days.
	}
\end{figure*}
 
\subsection{Categorical prediction of online items}
\label{subsec:category-prediction}

Here, we evaluate the categorization of online items outside Twitter (video categories, publisher reputation, IO state-backed agents) using solely the retweet cascades temporal information. 

\textbf{Setup.}
We follow the experimental setup outlined by \citet{kong2020describing}.
For \activeRT we predict the category of YouTube videos among four video categories (Entertainment, Gaming, Music and News\&Politics).
For \rncnix, we predict whether a publisher is controversial or reputable.
For \io, we distinguish among four user state-backed agent identities (Russia-, Iran-, Saudi Arabia-backed and Organic User).
We evaluate the prediction performance using the macro-F1 with the same 
50\%-50\% train-test split setup as \citep{kong2020describing} and employing 5\% of the training data for validation.
We compare our proposed IC-TH (with and without pre-trained weights) to state-of-the-art baselines TH~\citep{zuo2020transformer} and DMM~\citep{kong2020describing}. 
We also evaluate a version of IC-TH where retweet count data is excluded from the input --- i.e., we remove the missing event counts and show only the event times --- to evaluate the contribution of knowing how many events are missing.
Note the dash for IC-TH on \rncnix, as this dataset does not have any missing data; 
therefore IC-TH is identical to IC-TH w/o missing counts.

\textbf{Results.}
We show the category prediction results in \cref{tab:categorical_prediction}. 
We make three observations.
First, the classification task is more straightforward on the datasets \rncnix and \io than on \activeRT, as all algorithms obtain significantly higher macro-F1 scores.
Second, our pre-trained IC-TH outperforms all other baselines and IC-TH flavors.
Third, we estimate the performance impact of the three modeling contributions of pre-trained IC-TH over TH.
The new data representation introduced in \cref{subsec:missing_events} appears to have the highest impact (IC-TH w/o missing counts compared to TH).
Introducing missing counts and pre-training both lead to moderate performance increases.

\textbf{Effect of pre-training weights.}
To explore the benefits of pre-trained weights, we test the IC-TH model (with and without pre-training) on training sets of \activeRT of varying sizes (in percentage).
We repeat the train-test procedure 10 times for each training set size and report the mean and standard deviation of Macro-F1. 
All pre-training is performed on \activepretrain.
\cref{fig:f1_prop_train} depicts the performance increase with the size of the training dataset.
We see that the improvements from the pre-trained weights become notable and consistent when using more than $40\%$ of training data.

\section{Analyzing state-backed IO agents}
In this section, we perform an in-depth quantitative (\cref{subsec:user-clusters}) and qualitative analysis (\cref{subsec:qualitative-russia}) of the users in the \io dataset, studying both the temporal dynamics and the content of the tweets.

\subsection{Identify clusters of state-backed accounts}
\label{subsec:user-clusters}
\textbf{Method.}
We leverage IC-TH fine-tuned on \activepretrain
to derive cascade group embeddings for users --- a cascade group corresponds to each user and contains all the retweet cascades in which the user is involved. 
We visualize in \cref{fig:twitter_takedown} the user embeddings from \io with t-SNE. 
The figure shows a clear separation between the removed state-backed agents from the three countries and the organic users who were not removed.
This is not surprising given the macro-F1 score of $0.987$ obtained by the pre-trained IC-TH on \io (see \cref{tab:categorical_prediction}).
However, clear divisions also exist within each country group --- e.g., at least three clusters can be observed for Russia-linked users.
Next, we investigate whether the users in the same clusters have more in common than temporal tweeting patterns.

\begin{table}[tbp]
	\centering
	\caption{
		Mean $\pm$ standard deviation of Jaccard similarity in hashtag usage between pairs of users in different clusters (inter-cluster line) and within the same cluster (intra-cluster line) for users affiliated with each country.
		Higher scores indicate more similarity.
	}
\begin{tabular}{lrrr}
		\toprule
		 & Russia ($k=3$) & \multicolumn{1}{p{1.7cm}}{Saudi Arabia\newline($k=3$)} & Iran ($k=4$)\\
		\midrule
		Inter-cluster sim. & $0.064_{\pm 0.0183}$ & $0.038_{\pm 0.0467}$ & $0.034_{\pm 0.0171}$ \\
		Intra-cluster sim. & $0.133_{\pm 0.0289}$ & $0.260_{\pm 0.2390}$ & $0.089_{\pm 0.0405}$ \\
		\bottomrule
	\end{tabular}
\label{tab:hashtag_similarity}
	\vspace{-2mm}
\end{table} \textbf{Assess content similarity of clusters.}
We first quantitatively measure the pairwise similarities of users within clusters. 
For state-backed removed users, we apply the k-means clustering on the t-SNE space with the $k$ number manually selected to match the cluster numbers in \cref{fig:twitter_takedown}: $3$ for Russia, 3 for Saudi Arabia and 4 for Iran.
We then build the sets of hashtags used by each user and compute the Jaccard similarity scores between all pairs of users. 
We show in \cref{tab:hashtag_similarity} the average similarity scores and standard deviations of inter-cluster and intra-cluster user pairs. 
The scores consistently indicate higher similarity among users within the same cluster for all countries.
This indicates that users placed in the same cluster based on their temporal tweeting patterns (captured by IC-TH) exhibit similar hashtag usage.
In other words, the identified clusters appear coherent both temporally and content-wise. 

\subsection{Investigation of Russia-backed Information Operations strategies}
\label{subsec:qualitative-russia}
Here we analyze the agenda-setting episodes for each cluster and investigate for signs of possible coordination across clusters by checking whether they post similar messages with increased frequency on the same day.

First, we map the temporal posting patterns of the identified Russia-backed clusters (i.e., number of tweets per day) to determine the days of increased activity within each cluster.
We identify the ``spikes'' --- the days characterized by increased frequency of tweets from the users in each particular cluster --- and sample and analyze their tweets' content at peak times. 
For example, in \cref{fig:russia_op}, we observe a distinct spike in activity among \textit{Cluster 2} and \textit{Cluster 3} on 18-19 July 2014. 
On 17 July 2014, Malaysia Airlines flight MH17 was shot down using an anti-aircraft system belonging to the Russian military, killing all 298 passengers and crew on board. 
We see from the IO tweets data that on the next day (18 July), users from both \textit{Clusters 2 and 3} released an avalanche of accusatory tweets in an attempt to shift blame towards Ukraine. 
The content of the Russian-language tweets included hashtags such as \textit{\#KievShotDowntheBoeing}, \textit{\#KievProvocation} and \textit{\#KievTelltheTruth}. 
The narratives alleged Ukraine's responsibility for committing the attack.
They also drew conclusions regarding Ukraine's future as an independent state. 
These narratives were aligned with Russia's strategic geopolitical interests --- shifting the blame to Ukraine and delegitimizing the Ukrainian government allowed them to justify their military intervention in the region. 

Interestingly, while we can observe instances of coordinated activity across clusters like the one above, we can also see that each cluster exhibits unique information-spreading patterns. 
After the downing of MH17, \textit{Cluster 2} continued tweeting Russian news and messages of support for the Russian government in the Russian language. 
In contrast, users in \textit{Cluster 3} have switched to English in an apparent effort to target a new audience on the platform. 
Some of these accounts posted inspirational quotes and engaged in popular culture discussions marked by common hashtags such as \textit{\#music} and \textit{\#usa}. 
Others engaged in conversations with other users, providing relationship advice and even consolation. 
Then on 11/09/2014, all of these accounts started pushing information using the hashtag \textit{\#columbianchemicals} --- a known hoax claiming an explosion at a chemical plant in Centerville, Louisiana in 2014 \cite{delwiche2019computational}.
\textit{\#columbianchemicals} was not the only issue discussed in the cluster --- other parts were dedicated to spreading false narratives about Ukraine, including negative portrayals of Ukraine's Revolution of Dignity and snapshots of a happy civilian life in the Russian-occupied Crimea. 
Another slightly smaller fraction of these conversations was dedicated to discussing the Russian soul and the ``indomitable Russian spirit'' that were often linked to Christianity and Russia's much-celebrated victory in World War II. 
Thus, \textit{Cluster 3} captures a broad repertoire of Russia's attempts at foreign and domestic influence perpetrated by seemingly ``ordinary'' accounts.

Unlike \textit{Cluster 3}, the tweets in \textit{Clusters 1 and 2} were exclusively in the Russian language. 
\textit{Cluster 1} contains retweets of news reported by the Russian mainstream media. 
Most of the tweets from this cluster have a notably patriotic framing, praising the Russian president Vladimir Putin or Russia's achievements in the automotive industry and sports. 
On certain days, users from this cluster would participate in coordinated campaigns, such as the \textit{\#MadeinRussia} campaign that aimed to justify Russia's import substitution policy. 
Following the official announcement, these accounts shared information on various Russian-made goods and praised their advantages compared to their sanctioned analogs.

\textit{Cluster 2} is thematically similar to \textit{Cluster 1} but has a distinct prevalence of regional (rather than national) news with a mix of conservative sentiment. 
It has been activated to promote pro-Russian hashtags that mimic public sentiments, such as \textit{\#LavrovBeStrong} on 21/03/2015 or \textit{\#TurkeyAggressor} on 24/11/2015. 

Overall, the temporal analysis of coordinated behavior within clusters suggests they might represent separate organizational units, such as troll farms. 
This is a plausible scenario in which each organizational unit/troll farm would be engaged in a separate set of activities that constitute Russia's information operations for domestic or international audiences.
However, as we can observe in \cref{fig:russia_op}, in the case of significant political events, they can also join their forces and participate in coordinated campaigns across clusters.

\section{Conclusion}
This paper proposes the Interval-censored Transformer Hawkes (IC-TH), an extension to the Transformer Hawkes model that accounts for both observer event times and counts of missing events between observed events.
We start by observing that most Twitter feeds are sampled-down, and tweets are missing at varying rates.
We show that the number of missing tweets can be reconstructed using the `$retweeted\_count$' information embedded in each retweet.
We propose a novel data embedding scheme that accounts for both observed and missing data, a new log-likelihood function for training IC-TH and a contrastive learning approach for training it with large volumes of unlabeled data.

We test IC-TH on three large datasets with two learning tasks: future popularity prediction and online item categorization.
We show that IC-TH outperforms the current state-of-the-art baselines on all datasets and tasks: generative and neural network-based, interval-censored, and event time-based.
Most importantly, we show that the category of YouTube videos, the reputability of news publishers and the status of state-backed actors for Twitter users can be predicted using solely the timing of the retweet cascades associated with these items.

\textbf{IC-TH for IO detection.}
As validation suggests, grouping users solely based on their temporal interaction patterns represented by retweet cascades allows researchers to identify clusters of users engaged in information operations. 
Moreover, the clusters' granularity might be sufficient in many cases to identify each distinct information operation and its textual and hypertextual features. 
Based on these preliminary findings, the interval-censored Transformer Hawkes might be used to identify coordinated inauthentic behavior in other datasets independent of their content, making it a potentially helpful tool for combatting foreign interference on social technology platforms. 

\subsection*{Acknowledgments}
This work was supported by the Commonwealth of Australia (represented by the Defence Science and Technology Group) through a Defence Science Partnerships Agreement.
This research used resources and services from the National Computational Infrastructure (NCI), which is supported by the Australian Government. 

\bibliographystyle{ACM-Reference-Format}

\end{document}